\documentclass[a4paper]{jpconf}
\usepackage{graphicx}

\usepackage{latexsym}
\usepackage{amsmath,,calrsfs}
\usepackage{amsfonts}
\usepackage{amssymb}
\usepackage{amscd}
\usepackage{fancybox}
\usepackage{cite}
\usepackage{amsmath,amsfonts,amsbsy}
\usepackage{pstricks,pst-node}
\usepackage[small,bf,hang]{caption2}
\usepackage{graphicx}
\usepackage{epsfig}
\usepackage{psfrag}

\usepackage{float}

\psset{unit=1.3cm,linewidth=.5pt,radius=.2}  

\usepackage{multirow}                     
\usepackage{float}                          
\usepackage{lscape}                         
\usepackage{bm}

\newcommand{\beq}{\begin{equation}}
\newcommand{\eeq}{\end{equation}}
\newcommand{\ei}{\end{itemize}}
\newcommand{\bt}{\begin{tabular}}
\newcommand{\et}{\end{tabular}}
\newcommand{\bc}{\begin{center}}
\newcommand{\ec}{\end{center}}

\newcommand{\be}{\begin{equation}}
\newcommand{\ee}{\end{equation}}
\newcommand{\bea}{\begin{eqnarray}}
\newcommand{\eea}{\end{eqnarray}}
\newcommand{\ba}{\begin{array}}
\newcommand{\ea}{\end{array}}

\def\bbox{{\,\lower0.9pt\vbox{\hrule \hbox{\vrule height 0.2 cm
\hskip 0.2 cm \vrule height 0.2 cm}\hrule}\,}}
\newcommand{\dsl}{\pa \kern-0.5em /}

\makeatletter \@addtoreset{equation}{section} \makeatother

\def\slashchar#1{\setbox0=\hbox{$#1$}           
   \dimen0=\wd0                                 
   \setbox1=\hbox{/} \dimen1=\wd1               
   \ifdim\dimen0>\dimen1                        
      \rlap{\hbox to \dimen0{\hfil/\hfil}}      
      #1                                        
   \else                                        
      \rlap{\hbox to \dimen1{\hfil$#1$\hfil}}   
      /                                         
   \fi}

\begin{document}
\rightline{ UG-02/11 , UUITP-02/11, KUL-TF-02/11 }

\title{Brane solutions and integrability: a status report}

\author{W. Chemissany ${}^1$, P. Fr\'e ${}^2$, J.  Rosseel ${}^3$, A. S. Sorin ${}^4$, M. Trigiante ${}^5$, and  T. Van Riet ${}^6$}

\address{${}^1$ Afdeling Theoretische Fysica, Katholieke Universiteit Leuven Celestijnenlaan 200D bus 2415, 3001 Heverlee, Belgium}

\address{${}^2$ Dipartimento di Fisica Teorica, Universit\'a di
Torino, 
$\&$ INFN - Sezione di Torino\,, via P. Giuria 1, I-10125 Torino,
Italy}

\address{${}^3$ Centre for Theoretical Physics, University of Groningen,
    Nijenborgh 4, 9747 AG Groningen, The Netherlands}

\address{${}^4$ Bogoliubov Laboratory of Theoretical
Physics, Joint Institute for Nuclear Research, 141980 Dubna, Moscow Region,
Russia}
\address{${}^5$ Dipartimento di Fisica Politecnico di Torino,
C.so Duca degli Abruzzi, 24, I-10129 Torino, Italy}

\address{${}^6$Institutionen f\"{o}r Fysik och Astronomi, Box 803, SE-751 08 Uppsala, Sweden}

\ead{wissam@itf.fys.kuleuven.be, fre@to.infn.it, j.rosseel@rug.nl, sorin@theor.jinr.ru, mario.trigiante@polito.it,thomas.vanriet@fysast.uu.se }

\begin{abstract}

 We review the status of the integrability and solvability of the geodesics equations of motion on symmetric coset spaces that appear as sigma models of supergravity theories when reduced over respectively the timelike and spacelike direction. Such geodesic curves describe respectively timelike and spacelike brane solutions. We emphasize the applications to black holes.
\end{abstract}

The construction of black hole solutions using sigma model techniques becomes very beneficial in supergravity theories with enough symmetries as first employed in \cite{Breitenlohner:1987dg}. This has later been extended to more general stationary p-brane solutions \cite{Gal'tsov:1998yu} and even time-dependent solutions (spacelike branes)\cite{Fre:2003ep,Fre':2007hd,Fre:2009zz,Fre:2005bs}, see \cite{Bergshoeff:2008be} for the general framework. The sigma model technique reduces the problem to the one of geodesic motion on the appropriate moduli spaces. In the case of time-dependent solutions the moduli space is a Riemannian coset space $G/H$, where $H$ is compact and in the stationary case a non-Riemannian coset space, $G/H^*$, where $H^*$ is non-compact. In this status report we review the recent insights of \cite{Chemissany:2009af, Chemissany:2009hq, Chemissany:2010zp, Fre:2009et, Chemissany:2010ay} on the integrability and solvability of the geodesic equations of motion when the coset moduli spaces are symmetric. In these works we established the following results
\begin{itemize}
 \item 
We made use of the group-theoretical structure of the target space to prove that the second 
order geodesic equations are integrable in the sense of Liouville \cite{Chemissany:2010zp}, by explicitly constructing the correct number of conserved Poisson-commuting constants of motion from the underlying Poisson structure of the dual Borel algebras associated with $G/H^*$ \cite{Fre:2009et}.

\item We presented an integration method by the construction of a Lax algorithm that integrates the second order equations in one step \cite{Chemissany:2010zp}, improving on the previously constructed two-step algorithms \cite{Fre:2009et, Fre:2009dg, Chemissany:2009af, Chemissany:2009hq } (see \cite{Fre':2007hd,Fre:2009zz,Fre:2005bs} for the case of Riemannian 
manifolds).
 
\item In the case of black holes we showed that Liouville integrability implies that there always exists a description in terms of a (fake) superpotential $\mathcal{W},$ in the effective potential approach (see e.g.\cite{Ceresole:2007wx,Andrianopoli:2009je,Chemissany:2010zp,Andrianopoli:2010bj,Bossard:2009we,Perz:2008kh} and refs therein). The correspondence between the effective potential and the sigma model approach was studied in depth in \cite{Chemissany:2010ay}  by establishing an interesting generalization of Toda molecule equations.

\end{itemize}

In analogy with the Riemannian case \cite{Fre:2005bs},  the geodesic equations on a symmetric space with indefinite signature can be recast into the Lax form \cite{Chemissany:2009hq, Fre:2009et}
\begin{equation}\label{laxgeo}\dot{ L}(\tau)=[L(\tau),W(\tau)],\end{equation}
where the dot denotes the differentiation w.r.t $\tau.$  $L$ and $W$ are respectively the Lax operator and connection, being defined in terms of the pull-back  $\Omega$ on the geodesic of the left-invariant 1-form on $G/H^*,$ i.e., in terms of a coset representative $\mathbb{L},$
\begin{equation}\Omega=\mathbb{L}^{-1}\dot{\mathbb{L}}=L+W.\end{equation}
The corresponding geodesic action can thus be written as 
$ S\sim \int \textrm{d} \tau\, \textrm{Tr}(LL).$ Choosing the solvable gauge\footnote{The solvable parametrization on $G/H^*$ amounts to defining a coset representative $\mathbb{L} \in G/H^*, $ namely, 
$\mathbb{L}(\phi)=\textrm{exp}{(\Phi^{i}T_{i})},$ where $T_{i}$'s  are the generators of the solvable algebra.}, one has
\begin{eqnarray}W&=& L_>-L_<,\end{eqnarray} where $L_{<(>)}$ denotes the upper-triangular (resp. lower-triangular) part of $L.$ The Lax algorithm that we have devised to solve (\ref{laxgeo}) is summarized by the following formula for the solution of the coset representative \cite{Chemissany:2010zp}
\begin{eqnarray}
\label{solutioncosetrepr}
\left(\mathbb{L}(\tau)^{-1}\right)_{ij} &\equiv&\frac{1}{\sqrt{\mathfrak{D}_i(\mathcal{C})\mathfrak{D}_{i-1}(\mathcal{C})}}\mathrm{Det}\left(\begin{array}{cccc} \mathcal{C}_{1,1}(\tau)&\dots
&\mathcal{C}_{1,i-1}(\tau)&
(\mathcal{C}(\tau)\mathbb{L}(0)^{-1})_{1,j}\\
\vdots&\vdots&\vdots&\vdots\\
\mathcal{C}_{i,1}(\tau)&\dots &
\mathcal{C}_{i,i-1}(\tau)& (\mathcal{C}(\tau)\mathbb{L}(0)^{-1})_{i,j}\\
\end{array}\right)\,,\nonumber
\end{eqnarray}
and we have defined
  \begin{equation}\label{cijN}
    \mathcal{C}(\tau)\, := \rme^{-2\, \tau \, L_0}\,,
\end{equation}
\begin{equation}\label{DktN}
    \mathfrak{D}_{i}(\mathcal{C}) \, := \, \mbox{Det} \, \left ( \begin{array}{ccc}
    \mathcal{C}_{1,1}(\tau) & \dots & \mathcal{C}_{1,i}(\tau)\\
    \vdots & \vdots & \vdots \\
    \mathcal{C}_{i,1}(\tau) & \dots & \mathcal{C}_{i,i}(\tau)
    \end{array}\right)  \, , \quad
    \mathfrak{D}_{0}(\tau):=1, \,
    \end{equation}
where $L_{0}$ and $\mathbb{L}(0)$ are, respectively, the initial values of the Lax matrix $L(\tau)$ and the coset representative $\mathbb{L}(\tau).$ The above equations generalize analogous equations written in \cite{Fre':2007hd,Fre:2009zz,Fre:2005bs} for the Lax operator in the case of Rimannian manifolds.

Having the Lax form does not yet imply the Liouville integrability of the system. By Liouville integrability we mean that there exist $n$ functionally independent constants of motion (Hamiltonians $\mathcal{H}_{i}$) satisfying $\{\mathcal{H}_{i},\mathcal{H}_{j}\}=0.$

Let us summarize the proof of Liouville integrability of the full \emph{second order} geodesic equations presented in detail in \cite{Chemissany:2010zp}. 
 Consider the Noether charge matrix
\begin{equation}Q=\mathbb{L}(\tau) L(\tau)\mathbb{L}(\tau)^{-1},\quad Q=Q^{A}T_{A},\quad L=Y^{A} T_{A},\end{equation}
where $T_{A}$ are the generators of the solvable algebra. One can verify the following Poisson brackets
\begin{equation}\label{pb}\{Y_{A},Y_{B}\}=-f_{AB}{}^{C}Y_{c},\qquad \{Q_{A},Q_{B}\}=f_{AB}{}^{C}Q_{C}.\end{equation} Note that $Y_{A}$ and $Q_{A}$ depend on the phase space variables of the Hamiltonian system. Denoting the dimension of the coset by $n$ and the dimension of the symplectic leaves by $2 h_{O},$ One can show that
\begin{eqnarray} 2 h_{O}&=& \textrm{rank}(f_{AB}{}^{C})\\
\# \textrm{Hamiltonians}&=&n-h_{O}\\
\# \textrm{Casimirs}&=&n-2 h_{O}\end{eqnarray}
If we denote by
\begin{equation}\mathcal{H}_{a}(Y_{A}),\quad a=1,\cdots, h_{O}; \qquad \mathcal{H}_{\ell}(Y_{A}),\quad \ell=1,\cdots,n-2 h_{O}\end{equation}
respectively the Hamiltonians on the leaves and the Casimirs, we can find the following $2(n-h_{O})$  constants of motion in involution (using identity (\ref{pb}))

\begin{equation}\mathcal{H}_{a}(Y_{A}),\qquad \mathcal{H}_{\ell}(Y_A),\qquad \mathcal{H}_{a}(Q_{A}),\qquad  \mathcal{H}_{\ell}(Q_A),\end{equation}
where the $\mathcal{H}(Q)$'s are obtained by replacing $Y_{A}$ by $Q_{A}$ in the $\mathcal{H}(Y)$'s.
One can demonstrate that each $\mathcal{H}_{\ell}(Q_{A})$ is a Casimir on its own and can thus be expressed as a function of $\mathcal{H}_{\ell}(Y_{A}).$ This therefore implies that the total number of Hamiltonians  in involution is $(n-h_{O})+h_{O}=n,$ the dimension of $G/H^*,$ thereby proving Liouville integrability. The explicit construction and general form of the Hamiltonians are discussed in \cite{Chemissany:2010zp} for the case of $\textrm{SL}(n,\mathbb{R}).$  Amongst the $\mathcal{H}(Y)$'s are certain number of $\emph{polynomial}$ conserved constants of motion which have a simple interpretation in the context of black holes. One of them corresponds to the extremality parameter $c.$ The remaining polynomial $\mathcal{H}(Y)$ keeps track of the regularity of the solution (they have to vanish for regular solutions). The Taub-NUT charge corresponds to one of the remaining non-polynomial $\mathcal{H}(Y),$ and electro-magnetic charges of the black hole solutions are functions of $\mathcal{H}(Q).$

This has been shown to imply that the Hamilton-Jacobi formalism, HJ, can be applied. This allows us, formally, to rewrite the momenta $P_{i}$ as functions of $\phi,$ the canonical coordinates (scalars parameterizing $G/H^* $), provided $J=\textrm{det}(\partial \mathcal{H}_{i}/\partial P_{j})\neq 0$ and conclude that (locally)
 \begin{equation} \dot{\phi}^i=g^{ij}(\phi) P_{j}(\phi,\mathcal{H}),\qquad P=d \mathcal{W},\end{equation}
which is nothing else than the standard HJ equations. In this sense the function $\mathcal{W}$ is the \emph{Hamilton's characteristic function}  that solves the HJ equations
\begin{equation} \mathcal{H}\left(\frac{\partial\mathcal{W}}{\partial\phi^{i}},\phi^i\right)=\frac{1}{2}\frac{\partial\mathcal{W}}{\partial\phi^{i}} g^{ij}(\phi)\frac{\partial\mathcal{W}}{\partial\phi^{j}}=c^{2},\end{equation}
where $c$ is the affine velocity along the geodesic. In problems related to black holes, these results solve an open question about the existence of a fake superpotential for black hole solutions. We have proven the (local) existence of a fake superpotential  for \emph{all} stationary, spherically symmetric, black hole solutions to symmetric supergravity theories. In \cite{Chemissany:2010zp} we illustrated this in an explicit example, $\textrm{SL}(3,\mathbb{R})$-model, in which we moreover have proven that the Jacobian $J$ is always non-vanishing in the subspace of regular solutions, thus ensuring the global existence of $\mathcal{W}.$

As an application to black hole solutions,  we have explicitly worked out in \cite{Chemissany:2010zp} the three findings mentioned in the introduction for the dilatonic black holes arising in Kaluza-Klein theories. Due to the simplicity of the dilatonic black hole model we are not able to understand the interpretation for the generic Hamiltonians as the model has only one non-polynomial $\mathcal{H}(Y)$, which is the Taub-NUT charge. Three of us are now trying to investigate the Hamiltonians in more involved models, such as the STU and $G_{2}$ models, which seems to make the physical interpretation more apparent.

Finally we want to end with a few interesting open problems that we would like to address in the future. First, it seems reasonable to extend the proof of Liouville integrability to the case of homogeneous non-symmetric spaces. Second, as already mentioned, it would be interesting to understand the physical interpretation of all constants of motion. As a further direction of investigation it might be useful to examine the role of the complete system of Hamiltonians in the context of radial quantization of black holes as a complete system of observables to uniquely define  the
quantum state.


\ack

W.C. is supported in part by the Natural Sciences and Engineering Research Council
(NSERC) of Canada. The work of M.T. is supported by the government grant PRIN 36
2007. T.V.R. is supported by the G\"oran Gustafsson Foundation.  The work of A.S. was partially
supported by the RFBR Grants. W.C. wishes to thank the organizers for a stimulating and inspiring atmosphere. W.C. and T.V.R. also thank the university of Granada for hospitality.


\section*{References}


\begin{thebibliography}{99}

\bibitem{Breitenlohner:1987dg}
P.~Breitenlohner, D.~Maison and G.~W. Gibbons,  {\em Four-dimensional black
  holes from Kaluza-Klein theories}, Commun. Math. Phys. {\bf 120} (1988)
295.

\bibitem{Gal'tsov:1998yu}
  D.~V.~Gal'tsov and O.~A.~Rytchkov,
  ``Generating branes via sigma-models,''
  Phys.\ Rev.\  D {\bf 58}, 122001 (1998)
  [arXiv:hep-th/9801160].
\bibitem{Fre:2003ep}
  P.~Fre, V.~Gili, F.~Gargiulo, A.~S.~Sorin, K.~Rulik and M.~Trigiante,
  ``Cosmological backgrounds of superstring theory and solvable algebras:
  Oxidation and branes,''
  Nucl.\ Phys.\  B {\bf 685}, 3 (2004)
  [arXiv:hep-th/0309237].
\bibitem{Fre':2007hd}
  P.~Fre' and A.~S.~Sorin,
  ``The arrow of time and the Weyl group: all supergravity billiards are
  integrable,''
  arXiv:0710.1059 [hep-th].
\bibitem{Fre:2009zz}
  P.~Fre and A.~S.~Sorin,
  ``The Weyl group and asymptotics: All supergravity billiards have a closed
  form general integral,''
  Nucl.\ Phys.\  B {\bf 815} (2009) 430.
\bibitem{Fre:2005bs}
P.~Fre and A.~Sorin,  {\em {Integrability of supergravity billiards and the
  generalized Toda lattice equation}}, Nucl. Phys. {\bf B733} (2006) 334--355.



\bibitem{Bergshoeff:2008be}
E.~Bergshoeff, W.~Chemissany, A.~Ploegh, M.~Trigiante and T.~Van~Riet,  {\em
  {Generating Geodesic Flows and Supergravity Solutions}}, Nucl. Phys. {\bf
  B812} (2009) 343--401.

\bibitem{Chemissany:2009hq}
W.~Chemissany, J.~Rosseel, M.~Trigiante and T.~Van~Riet,  {\em {The full
  integration of black hole solutions to symmetric supergravity theories}},
  Nucl. Phys. {\bf B830} (2010) 391--413.


\bibitem{Fre:2009et}
  P.~Fre and A.~S.~Sorin,
  ``Supergravity Black Holes and Billiards and Liouville integrable structure
  of dual Borel algebras,''
  JHEP {\bf 1003}, 066 (2010)
  [arXiv:0903.2559 [hep-th]].
 

\bibitem{Fre:2009dg}
  P.~Fre and A.~S.~Sorin,
  ``The Integration Algorithm for Nilpotent Orbits of $G/H^{*}$ Lax systems: for
  Extremal Black Holes,''
  arXiv:0903.3771 [hep-th].

\bibitem{Ceresole:2007wx}
  A.~Ceresole, G.~Dall'Agata,
  ``Flow Equations for Non-BPS Extremal Black Holes,''
  JHEP {\bf 0703 } (2007)  110.
  [hep-th/0702088].
\bibitem{Chemissany:2009af}
W.~Chemissany, P.~Fre and A.~S. Sorin,  {\em {The Integration Algorithm of Lax
  equation for both Generic Lax matrices and Generic Initial Conditions}},
  Nucl. Phys. {\bf B833} (2010) 220--225.
\bibitem{Chemissany:2010zp}
  W.~Chemissany, P.~Fre, J.~Rosseel, A.~S.~Sorin, M.~Trigiante and T.~Van Riet,
  ``Black holes in supergravity and integrability,''
  JHEP {\bf 1009}, 080 (2010)
[arXiv:1007.3209 [hep-th]].
\bibitem{Chemissany:2010ay}
  W.~Chemissany, J.~Rosseel and T.~Van Riet,
  ``Black holes as generalised Toda molecules,''
  Nucl.\ Phys.\  B {\bf 843}, 413 (2011)
  [arXiv:1009.1487 [hep-th]].
\bibitem{Perz:2008kh}
J.~Perz, P.~Smyth, T.~Van~Riet and B.~Vercnocke,  {\em {First-order flow
  equations for extremal and non-extremal black holes}}, JHEP {\bf 03} (2009)
  150.
\bibitem{Andrianopoli:2009je}
  L.~Andrianopoli, R.~D'Auria, E.~Orazi and M.~Trigiante,
  ``First Order Description of D=4 static Black Holes and the Hamilton-Jacobi
  equation,''
  Nucl.\ Phys.\  B {\bf 833}, 1 (2010)
  [arXiv:0905.3938 [hep-th]].
\bibitem{Andrianopoli:2010bj}
  L.~Andrianopoli, R.~D'Auria, S.~Ferrara and M.~Trigiante,
  ``Fake Superpotential for Large and Small Extremal Black Holes,''
  JHEP {\bf 1008}, 126 (2010)
  [arXiv:1002.4340 [hep-th]].
\bibitem{Bossard:2009we}
G.~Bossard, Y.~Michel and B.~Pioline,  {\em {Extremal black holes, nilpotent
  orbits and the true fake superpotential}}, JHEP {\bf 01} (2010) 038.

\end{thebibliography}
\end{document}